\documentstyle[sprocl,epsfig]{article}

\def\la{\mathrel{\mathchoice {\vcenter{\offinterlineskip\halign{\hfil
$\displaystyle##$\hfil\cr<\cr\sim\cr}}}
{\vcenter{\offinterlineskip\halign{\hfil$\textstyle##$\hfil\cr<\cr\sim\cr}}}
{\vcenter{\offinterlineskip\halign{\hfil$\scriptstyle##$\hfil\cr<\cr\sim\cr}}}
{\vcenter{\offinterlineskip\halign{\hfil$\scriptscriptstyle##$\hfil\cr<\cr
\sim\cr}}}}}

\def\ga{\mathrel{\mathchoice {\vcenter{\offinterlineskip\halign{\hfil
$\displaystyle##$\hfil\cr>\cr\sim\cr}}}
{\vcenter{\offinterlineskip\halign{\hfil$\textstyle##$\hfil\cr>\cr\sim\cr}}}
{\vcenter{\offinterlineskip\halign{\hfil$\scriptstyle##$\hfil\cr>\cr\sim\cr}}}
{\vcenter{\offinterlineskip\halign{\hfil$\scriptscriptstyle##$\hfil\cr>\cr
\sim\cr}}}}}

\begin{document}

\title{FIRST RESULTS FROM THE \\ KASCADE AIR SHOWER 
EXPERIMENT\footnote{Talk given at the
Second Meeting on New Worlds in Astroparticle Physics, University of the
Algarve, Faro, Portugal, September 1998}}

\author{
K.-H.~Kampert$^{b,a,}\footnote{E-Mail: \tt{kampert@ik1.fzk.de}}$,
T.~Antoni$^{a}$, 
W.D.~Apel$^{a}$, 
K.~Bekk$^{a}$, 
K.~Bernl\"ohr$^{a}$,
E.~Bollmann$^{a}$,
H.~Bozdog$^{c}$,
I.M.~Brancus$^{c}$,
A.~Chilingarian$^{d}$,
K.~Daumiller$^{b}$, 
P.~Doll$^{a}$, 
J.~Engler$^{a}$, 
M.~Fessler$^{a}$, 
H.J.~Gils$^{a}$,
R.~Glasstetter$^{b}$, 
R.~Haeusler$^{a}$, 
W.~Hafemann$^{a}$, 
A.~Haungs$^{a}$, 
D.~Heck$^{a}$, 
T.~Holst$^{a}$, 
J.~H\"orandel$^{b}$, 
J.~Kempa$^{e}$,
H.O.~Klages$^{a}$, 
J.~Knapp$^{b}$, 
H.J.~Mathes$^{a}$, 
H.J.~Mayer$^{a}$, 
J.~Milke$^{a}$, 
D.~M\"uhlenberg$^{a}$, 
J.~Oehlschl\"ager$^{a}$, 
M.~Petcu$^{c}$, 
H.~Rebel$^{a}$, 
M.~Risse$^{a}$, 
M.~Roth$^{a}$, 
G.~Schatz$^{a}$, 
H.~Schieler$^{a}$, 
F.K.~Schmidt$^{b}$, 
T.~Thouw$^{a}$, 
H.~Ulrich$^{a}$,
B.~Vulpescu$^{c}$, 
J.~Weber$^{a}$, 
J.~Wentz$^{a}$, 
T.~Wibig$^{e}$, 
T.~Wiegert$^{a}$, 
D.~Wochele$^{a}$, 
J.~Wochele$^{a}$, 
J.~Zabierowski$^{e}$\\ 
{\bf --- KASCADE Collaboration ---}\\[1ex]}

\address{
$^{a}$ Institut\ f\"ur Kernphysik, Forschungszentrum Karlsruhe, and\\
$^{b}$ Inst.\ f\"ur Exp. Kernphysik,
           Universit\"at Karlsruhe, D-76021 Karlsruhe, Germany,\\
$^{c}$ Inst.\ of Physics and Nuclear Engineering,
           RO-7690 Bucharest, Romania, \\
$^{d}$ Cosmic Ray Division, Yerevan Physics Inst., 
           Yerevan 36, Armenia, \\
$^{e}$ Inst.\ for Nuclear Studies and Dept.\ of
          Exp. Physics, University of Lodz,
          PL-90950 Lodz, Poland
}

\maketitle\abstracts{ The extensive air shower (EAS) experiment 
KASCADE has started data taking at the laboratory site of the 
Forschungszentrum Karlsruhe.  The major goal is to determine the 
elemental composition of the primary cosmic ray particles in the 
energy range around and above the knee at $E_{\rm k} \approx 5 \cdot 
10^{15}$ eV. Here, we shall discuss some results on tests of hadronic 
interaction models required for air shower simulations, present the 
`knee' in particle size distributions of electrons, muons, and 
hadrons, and finally discuss preliminary results on the elemental 
composition which are based on event-by-event measurements of the 
muon/electron ratio.  The great advantage over previous analyses which 
were based purely on {\em average} numbers of samples of events is 
emphasized.  The KASCADE data, analyzed in variety of different 
approaches, favor an increasingly heavier composition above the knee.  
}

\section{Introduction}

The origin and acceleration mechanism of ultra-high energy cosmic rays 
have been subject to debate for several decades.  Mainly for reasons 
of the required power the dominant acceleration sites are generally 
believed to be supernova remnants in the Sedov phase.  Naturally, this 
leads to a power law spectrum as is observed experimentally.  Detailed 
examination suggests that this process is limited to $E/Z \la 
10^{15}$\,eV. Curiously, the CR spectrum steepens at approx.\ $5 
\times 10^{15}$\,eV, indicating that the `knee' may be related to the 
upper limit of acceleration.  A change in the CR propagation with 
decreasing galactic containment has also been considered.  A key 
observable for understanding the origin of the knee and distinguishing 
the SN acceleration model from other proposed mechanisms, is given by 
the mass composition of CR particles and by possible variations across 
the knee (see e.g.\ Ref.\cite{berezhko98}).  Unfortunately, beyond the 
knee little is known about the CR's other than their energy spectrum 
\cite{watson98}.  The low flux of particles ($\sim 1$ 
m$^{-2}\cdot$year$^{-1}$ above the knee) puts strong demands on the 
collection power of the experiments, such as can only be achieved by 
extended air shower (EAS) arrays at ground level.  Sampling detector 
systems with typical coverages of less than one percent can be used 
for registration of such EAS. However, this indirect method of 
detection bears a number of serious difficulties in the interpretation 
of the data and requires detailed modeling of the air shower 
development and detector responses.  It is well known, that a number 
of characteristics of EAS depends on the energy per nucleon of the 
primary nucleus, notably the ratio of electron to muon numbers, the 
energy of the hadrons in the shower, the shapes of the lateral 
distributions of the various components of the shower, etc.\ The basic 
concept of the KASCADE experiment is to measure a large number of 
these parameters in each individual event in order to verify the 
consistency of EAS simulations and to determine both the energy and mass 
of the primary particles from the experimental data on a reliable 
basis.

\begin{figure}[tb]
\begin{center}
\epsfig{file=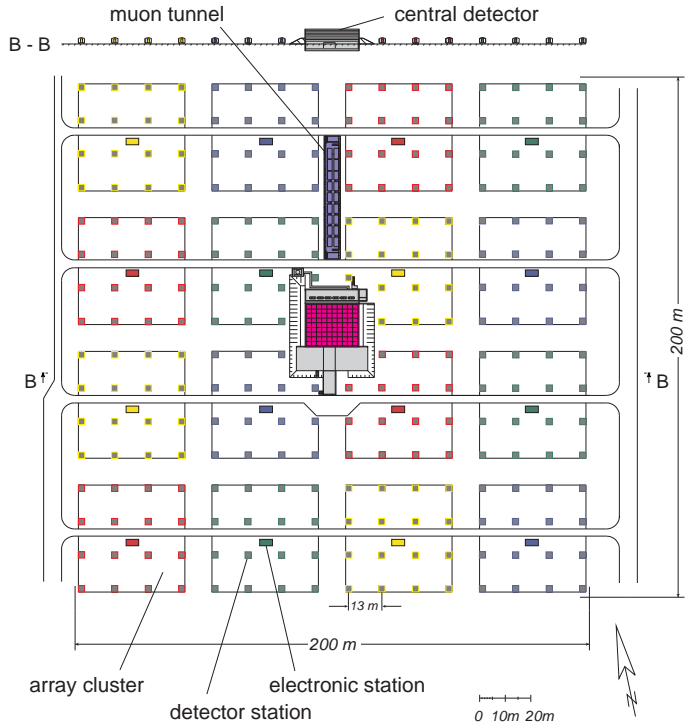,width=9cm}
\caption{Schematic layout of the KASCADE experiment.}
\label{fig:kascade-layout}
\end{center}
\end{figure}

\section{Layout and Status of the Experiment}

KASCADE (\underline{Ka}rlsruhe \underline{S}hower \underline{C}ore and 
\underline{A}rray \underline{De}tector) is located on the laboratory 
site of the Forschungszentrum Karlsruhe, Germany (at $8^{\circ}$ E, 
$49^{\circ}$ N, 110 m a.s.l.).  In brief, it consists of three major 
components (see Fig.\,\ref{fig:kascade-layout});

\begin{enumerate}
    \item A scintillator array comprising 252 detector stations of 
    electron and muon counters arranged on a grid of $200 \times 200$ 
    m$^{2}$.

    \item A central detector system (320 m$^{2}$) with a highly-segmented
    hadronic calorimeter, read out by 40,000 channels of warm 
    liquid ionization chambers, a scintillator trigger plane in the 
    third layer, and at the very bottom, 2 layers of positional 
    sensitive MWPC's for muon tracking at $E_{\mu} \ge 2$ GeV.

    \item A $48 \times 5.4$ m$^{2}$ tunnel housing three layers of 
    positional sensitive limited streamer tubes for muon tracking at 
    $E_{\mu} \ge 0.8$ GeV. This part of the experiment will be fully 
    operational by summer 1999 and has not been available for the data 
    analysis presented below.

\end{enumerate}

More details about the experiment can be found in Refs.\ 
\cite{proposal,klages-96}.  First data taking has started in late 1995 
with large parts of the experiment in stand-alone mode.  Correlated 
data are being collected with some parts of the experiment since April 
1996 and with its full set-up since early 1998.  By the time of 
writing this report, more than 160 Mio.\ events have been collected in 
a very stable mode with a trigger threshold corresponding to $E \sim 4 
\cdot 10^{14}$ eV.

\begin{figure}[p]
\centerline{\epsfig{file=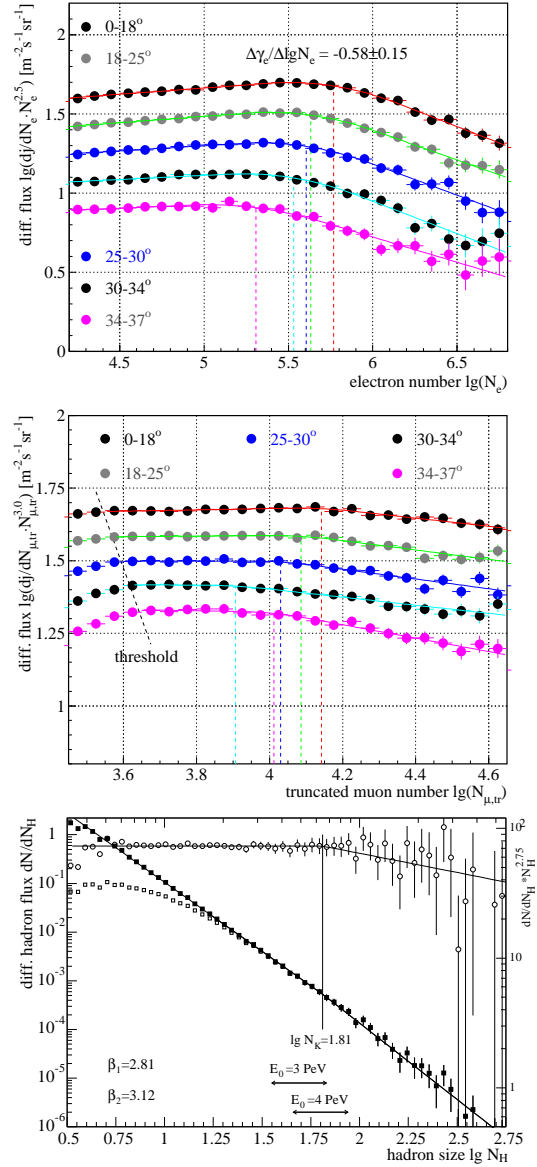,width=7.0cm}}
\caption[xx]{Electron, muon, and hadron size distributions.}
\label{fig:size-distributions}
\end{figure}

\section{First Results}

The central detector system is a unique feature of KASCADE and opens a 
new window to EAS. Firstly, high energy hadrons ($E_{\rm h} \ga 50$ 
GeV) detected within the core of EAS provide important information 
about the quality of hadronic interaction models used for EAS 
simulations.  Secondly, observables extracted from the shower core 
exhibit a strong sensitivity to the mass of primary particle, 
particularly in the energy range below the knee.  Detailed 
investigations have been performed to compare CORSIKA simulations 
\cite{corsika} using different hadronic interaction models (QGSJET, 
VENUS, SIBYLL) with experimental data \cite{hoerandel-98}.  Generally, 
the SIBYLL model, traditionally being used in EAS simulations, results 
in a very poor description of the experimental data.  This is most 
obviously exemplified in the energy spectrum and spatial distribution 
of high energy hadrons \cite{hoerandel-98}.  The observed discrepancy 
may only partly be attributed to a deficit in the predicted muon 
number by SIBYLL. On the other hand, VENUS and particularly GQSJET are 
found to describe most experimental data reasonably well when assuming 
an increasingly heavier composition with increasing energy.  
Significant deviations start to show up only for muon sizes 
corresponding to primary energies larger than $\sim 10^{16}$ eV. As 
the present analysis is limited to the knee region, this is of no 
major concern for the results presented below.  Nevertheless, a new 
and improved model based on VENUS and QGSJET is expected to become 
available soon \cite{nexus}.

Shower size spectra may be considered {\em direct} 
observables of EAS, because they are extracted from the experimental 
data, basically independent from any simulation.  The experimental 
set-up of KASCADE allows independent measurements of all three shower 
components, i.e.\ electrons, muons, and hadrons.  After determination 
of the shower axis from the scintillator array, the different shower 
sizes are calculated by fitting the measured respective distributions 
of lateral particle density to NKG-like parametrisations and 
calculating their two-dimensional integral in appropriate regions 
(electrons: 0-$\infty$, muons: 40-200 m, hadrons: 0-24 m).  This 
yields the electron size $N_{e}$, truncated muon size $N^{\rm 
tr}_{\mu}$, and hadron size $N_{\rm h}$, respectively.  The results 
are shown in Fig.\ 2.  Electron and muon size distributions are -- 
because of superior statistics -- given for different bins of zenith 
angles.  In each of these distributions, a knee is clearly visible.  
The variation of the knee position with zenith angle is found to be in 
accordance with simulations and translates into an attenuation length 
of $230 \pm 37$ g/cm$^{2}$ and $488 \pm 65$ g/cm$^{2}$ for electrons 
and muons, respectively.  Averaged over zenith angle, the
electron, muon, and hadron size indices are measured to be 
$\gamma_{e} \simeq -2.43 \pm 0.10$, $-2.88 \pm 0.11$, $\gamma_{\mu} = 
-2.99 \pm 0.11$, $-3.16 \pm 0.20$, $\gamma_{h} = -2.81 \pm 0.04$, 
$-3.12 \pm 0.11$ below and above the knee, respectively.  The indices 
of the shower size distributions $dJ/dN_{x}$ are related to the 
primary energy spectrum by the relation $\frac{dJ}{dE_{0}} = 
\frac{dJ}{dN_{x}} \frac{dN_{x}}{dE_{0}}$.  Taking the second factor 
from CORSIKA simulations we then obtain $\gamma_{\rm prim} = 2.75 \pm 
0.11$, $3.04 \pm 0.16$ for the primary energy spectrum below and above 
the knee, respectively. The knee position is found at $E_{k} \simeq 4$ PeV.  We 
like to point out, that a consistent description for all particle 
species can only be obtained, if an {\em increasingly heavier} 
composition is assumed above the knee.

\begin{figure}[t]
\centerline{\epsfig{file=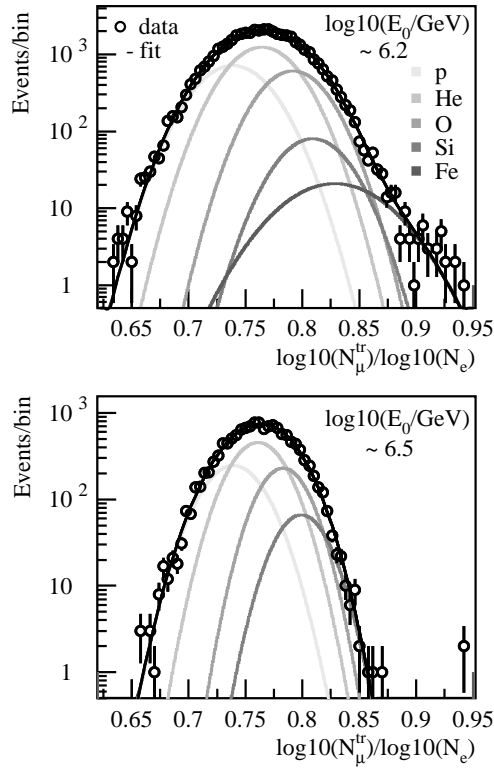,width=6.5cm}}
\caption[xx]{Distribution of the $\log N^{tr}_{\mu}/$ $\log N_{e}$ sizes.
The open circles represent the experimental data. The lines are fits 
to CORSIKA/QGSJET simulations for different primary particles as 
indicated. The sum of these simulated distributions is made to fit 
the experimental data.}
\end{figure}

\begin{figure}[t]
\centerline{\epsfig{file=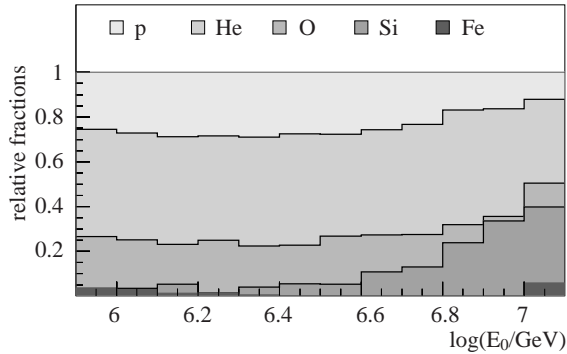,width=7.5cm}}
\caption[xx]{Extracted composition based on CORSIKA/QGSJET 
simulations (prelimary data). The relative fractions are visualized
by their respective areas.}
\end{figure}

A direct link to the mass of the primary particle is given 
by the muon to electron ratio which in KASCADE can be measured on an 
event-by-event basis.  An example of such an analysis is presented in 
Fig.\ 3 for two different muon-size bins, one below and one above the 
knee.  Results of CORSIKA simulations for different primary elements 
are shown by the grey lines \cite{weber}.  In order to extract the 
elemental composition at this energy (i.e.\ muon-size), the relative 
weights of the simulated distributions are fitted to the experimental 
data.  The sum of the distributions is shown by the black line, 
demonstrating the perfect fit to the data.  This is remarkable, 
because the left hand tail of the experimental data is matched 
perfectly by the proton simulations while the right hand tail is 
matched by the iron simulations.  Furthermore, the width of the 
fluctuations observed in the data seem to be inconsistent with a pure 
single mass component.  Evidently, such kind of analyses are superior 
to those, where only mean values of distributions can be compared.

Comparing the deconvolution of the two distributions in Fig.\ 3, one 
observes a significant change from a light to a medium-heavy dominated 
composition.  This change of the composition is summarized in Fig.\ 4.  
Here, the relative contributions are plotted as a function of the 
primary energy, with the latter being extracted from the muon size and 
taking into account the extracted mass composition.  Above 4 PeV 
($\log(E/{\rm GeV}) \simeq 6.6$) an increasing fraction of heavy 
primaries is observed.

\section{Summary and Conclusions}

Preliminary data of KASCADE have been presented.  For the first time, 
a knee is observed in the electron, muon, and hadron size spectra, 
indicating that the knee is an inherent feature of the primary energy 
spectrum.  The steep electron size spectrum above the knee can only be 
explained to be consistent with the muon size spectrum, if the 
composition above the knee gradually changes from light to heavy.  
Furthermore, the new quality of data allows to extract the 
muon-to-electron ratio on an event-by-event basis, thereby giving a 
direct link to the mass of the primary particle.  A deconvolution into 
different elemental groups again results in an increasing fraction of 
Si+Fe primaries above the knee.  The results are confirmed also by 
related analyses of KASCADE observables (lateral distributions, 
shower-core parameters, etc.).  Multiparameter analyses with more 
sophisticated classification schemes are under development in order to 
include all relevant observables simultaneously.  In parallel, 
hadronic event generators need further attention particularly at 
energies above $10^{16}$ eV but EAS data such as obtained from KASCADE 
may serve as vital input.

\end{document}